\documentclass[twocolumn,jcp,citeautoscript,superscriptaddress]{revtex4}
\usepackage{amsfonts,amsmath,amssymb}
\usepackage{graphicx,color}

\newcommand{\MCS}{\textnormal{MCS}}
\newcommand{\tn}[1]{\textnormal{#1}}

\newcommand{\ve}[1]{\mathbf{#1}}

\begin{document}

\title[Protein-protein interaction]
{Membrane-mediated protein-protein interaction: A Monte Carlo study}

\author{J\"org Neder}
\affiliation{Department of Physics, University of Konstanz, Germany}

\author{Beate West}
\affiliation{Department of Physics, University of Bielefeld, Germany}

\author{Peter Nielaba}
\affiliation{Department of Physics, University of Konstanz, Germany}

\author{Friederike Schmid}
\email{friederike.schmid@uni-mainz.de}
\affiliation{Institute of Physics, University of Mainz, Germany}

\begin{abstract}

We investigate membrane-mediated interactions between transmembrane proteins
using coarse-grained models. We compare the effective potential of
mean force (PMF) between two proteins, which are always aligned parallel to
the $z$-axis of the simulation box, with those PMFs obtained for proteins with
fluctuating orientations.  The PMFs are dominated by an oscillatory
packing-driven contribution and a smooth attractive hydrophobic mismatch
contribution, which vanishes if the hydrophobic length of the protein matches
the thickness of the membrane. If protein orientations are allowed to fluctuate,
the oscillations are greatly reduced compared to proteins with fixed orientation.
Furthermore, the hydrophobic mismatch interaction has a smaller range.
Finally, we compare the two-dimensional thickness profiles around
two proteins with the predictions from the elastic theory of two coupled
monolayers, and find them to be in very good agreement.

\end{abstract}

\maketitle

\section{Introduction} 

Biological membranes are essential components of all living organisms.  
Even though their basic building block is a fluid lipid bilayer, their 
properties and functionality crucially depend on the membrane proteins.
Proteins function as catalysts, transport and store other molecules,
provide mechanical support and immune protection, generate movement, transmit
nerve impulses, and control growth and differentiation \cite{bertym:02}.

Unfortunately, the components of real biomembranes are too diverse and complex
to obtain detailed and unambiguous information about interactions of
transmembrane segments with a lipid bilayer \cite{depkil:03}.  Moreover,
structural perturbations or transformations of the lipid bilayer are among the
most difficult processes to probe experimentally \cite{may:00}. Thus,
complementary theoretical approaches and computer simulations of membrane
systems of well-defined compositions are necessary to elucidate the role of the
lipid bilayer in processes like protein aggregation and function. 
In this context, the use of coarse-grained models has become popular
\cite{vot:09,muekat:06,des:09,sch:09}. Even with today's computing
resources, atomistic modeling of multicomponent lipid bilayers on length 
scales of several nanometer still remains a challenge. Moreover, 
coarse-grained model simulations give insight into generic properties 
and mechanisms in lipid membranes, which are difficult to extract 
from fully atomistic simulations of specific membranes.

In this paper we focus on lipid-mediated interactions between proteins or, more
generally, membrane inclusions.  We present simulational and theoretical
results on fluid membranes containing model transmembrane proteins, i.e.
proteins that span through the membrane.  Many theoretical models have been
developed to get a better understanding of the membrane-mediated interactions
between such proteins
\cite{mar:76,danpin:93,araber:96,mayben:00,lagzuc:00,bohkra:03,brabro:06,brabro:07}.
In these models, the inclusions were modeled as straight cylinders, which are
aligned with the bilayer normal. Only recently, computational studies with
coarse-grained models have been performed on the same problem, 
where both proteins in single-component bilayers \cite{schgui:08,demven:08,
wesbro:09} and the effect of cholesterol on protein-lipid and
lipid-mediated protein-protein interactions have been studied
\cite{demrod:10}. 
From these studies, the following
general picture has emerged: The lipid-mediated protein interactions can be
divided in short-range and long-range contributions. The short-range
contributions depend on the local structure of the lipid layer and reflect
packing and layering effects. The long-range interactions result from the
elastic distortion of the membrane due to the insertion of the proteins and can
be tuned, by tuning the length of the hydrophobic section of the protein
(''hydrophobic mismatch'' interaction). If the hydrophobic length of the
proteins matches the thickness of the membrane, they will vanish, otherwise
they tend to be attractive. In the literature, a number of other mechanisms
that would induce long-range lipid-mediated protein interactions have been
discussed \cite{gou:96, wei:01, reyill:07}, but for straight cylindrical 
proteins, the hydrophobic mismatch interaction seems be dominant.

The work presented here is based on a coarse-grained lipid model
\cite{lensch:05, schdue:07}, which reproduces the main phases
\cite{lensch:07,wessch:10:2, nedwes:10} and elastic properties
\cite{wesbro:09,nedwes:10} of DPPC bilayers.  In previous papers, this model
was used to study the lipid-mediated interactions between infinitely strong
straight cylinders with fixed orientation along the bilayer normal in
stressfree \cite{wesbro:09} and stressed membranes \cite{nedwes:10}. The rigid
restriction was motivated by the huge amount of related theoretical work (see
above). In reality, however, proteins may tilt in the membrane, their
orientation may fluctuate, and this affects the lipid-mediated protein
interactions. 

The present paper focuses on this effect. We study the lipid-mediated protein
interactions, using two variants of a coarse-grained protein model. The first
variant is the infinitely long straight cylinder studied earlier
\cite{wesbro:09,nedwes:10}, and the second variant consists of a cylinder of finite length with full
freedom to tilt. This allows to assess in detail the effect of
orientation fluctuations on the protein-protein interactions. In addition,
we characterize the profiles of membranes containing two proteins
and compare them with the theoretical prediction of an elastic theory.

Our paper is organized as follows: In the next section, we introduce the 
simulation models and method and briefly comment on the elastic theory
with which the simulation data are compared. The simulation results are
presented in Section III. We summarize and conclude in Section IV.
 
\section{Models and Methods} 

In the following a brief summary of the lipid model and the protein models used
in this work will be given.  In our study, we vary both the hydrophobic
mismatch of the proteins (i.e., the length of the hydrophobic section), and the
hydrophobic strength.  Since hydrophobic interaction does not arise from the
binding of nonpolar molecules to each other but from preventing polar solvent
molecules from achieving optimal hydrogen binding, the strength of the
hydrophobic interaction depends on the relative polarity of both the solute and
the solvent \cite{lancra:10}.  Experimentally the hydrophobicity of proteins
can therefore be tuned by changing amino acid residues with different
hydrophobicity of the protein \cite{depkil:03}. Alanine, e.g., is  less
hydrophobic than leucine \cite{wimwhi:96}.  Alternatively, changing the pH of
the solvent and thereby making priorly uncharged side-chains charged, will also
affect the hydrophobic interaction of the lipid bilayer and the proteins
\cite{bec:96}.

\subsection{Lipid Model}
\label{sec:simulation}

Lipid molecules are represented by chains consisting of one head bead
and six tail beads, and there are additional solvent beads
\cite{schdue:07}. Within the lipid chain, adjacent beads at a distance 
$r$ interact {\em via} a finite
extensible non-linear elastic (FENE) potential
\begin{equation} 
V_{\mathrm{FENE}}(r) = - \frac{1}{2} \epsilon_{\mathrm{FENE}} 
  (\Delta r_{\mathrm{max}})^2 
  \ln \left( 1 - \frac{(r - r_0)^2}{\Delta r_{\mathrm{max}}^2}
  \right), 
\end{equation}
with the spring constant $\epsilon_{\mathrm{FENE}}$, the equilibrium
bond length $r_0$, and the cutoff $\Delta r_{\mathrm{max}}$.
The angles $\theta$  between subsequent bonds in the lipid are
subject to a stiffness potential 
\begin{equation}
V_{\textnormal{BA}}(\theta) = \epsilon_{\textnormal{BA}} (1 + \cos \theta),
\end{equation}
with the stiffness parameter $\epsilon_{\textnormal{BA}}$.
Beads of type $i$ and $j$ which are not direct next neighbors in a chain 
interact {\em via} a truncated and shifted Lennard-Jones potential,
\begin{equation} 
V_{\textnormal{LJ}}(r/\sigma_{ij}) = \left\{ \begin{array}{cl} \epsilon
\left( \frac{\sigma_{ij}^{12}}{r^{12}} 
- 2 \frac{\sigma_{ij}^6}{r^6}\right) - V_{c,ij}, & \textrm{if}\  
r < r_{c,ij} \\ 0 & \textrm{otherwise.} \end{array}
\right. 
\label{equ:V_LJ}
\end{equation}
The offset $V_{c,ij}$ is chosen such that
$V_{\textnormal{LJ}}(r/\sigma_{ij})$ is continuous at the cutoff
$r_{c,ij}$. The parameter $\sigma_{ij} = (\sigma_i + \sigma_j)/2$ is
the arithmetic mean of the diameters $\sigma_i$ of the interaction
partners, and $r_{c,ij} = 1\,\sigma_{ij}$ for all partners $(ij)$
except $(tt)$ and $(ss)$: $r_{c,tt} = 2\,\sigma_{tt}$ and $r_{c,ss} =
0$. Hence tail beads attract one another, all other interactions are
repulsive, and solvent beads do not interact at all with each other.


We use the model parameters \cite{schdue:07} $\sigma_h = 1.1 \,\sigma_t$, $r_0
= 0.7\,\sigma_t$, $\Delta r_{\mathrm{max}} = 0.2 \,\sigma_t$,
$\epsilon_{\mathrm{FENE}} = 100\,\epsilon/\sigma_t^2$, and
$\epsilon_{\textnormal{BA}} = 4.7\,\epsilon$. At the pressure $P =
2.0\,\epsilon/\sigma_t^3$, the model reproduces the main phases of
phospholipids, {\em i.e.}, a high-temperature fluid $L_\alpha$ phase at
temperature $k_B T > k_B T_m \sim 1.2\,\epsilon$ and a low-temperature tilted
gel ($L_{\beta'}$) with an intermediate modulated ripple ($P_{\beta'}$) phase
\cite{lensch:07}.  The energy and length scales can be mapped to SI-units
\cite{wesbro:09} by matching the bilayer thickness or, alternatively, the area
per lipid and the temperature of the main transition to those of DPPC, giving
$1\,\sigma_t \sim 6\,\textrm{\AA}$ and $1\,\epsilon \sim 0.36 \times
10^{-20}\,\mathrm{J}$.  The elastic properties of the membranes in the fluid
state were then also found to be comparable to those of DPPC membranes
\cite{wesbro:09}.

\begin{figure}[ht]
\begin{center}
\begin{minipage}{\columnwidth}
\begin{minipage}{0.45\textwidth}
\includegraphics[width=\textwidth]{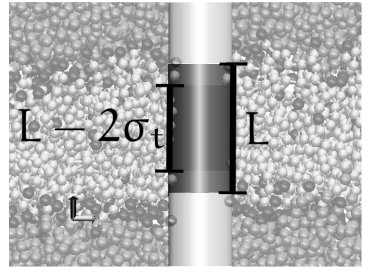}
\end{minipage}
\hfill
\begin{minipage}{0.45\textwidth}
\includegraphics[width=\textwidth]{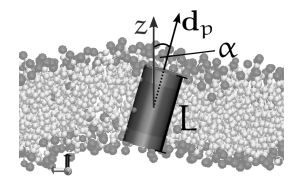}
\end{minipage}
\end{minipage}
\end{center}
\caption[]{Snapshots of the protein models considered in this work.
Left: Infinite cylinder with fixed orientation. Right: Finite tiltable cylinder.
$L$ denotes the total length of the hydrophobic section. The dark shaded band
in the middle indicates the region where the hydrophobic attractive interaction 
with lipid tails reaches full strength, the two adjacend light shaded bands
show the the region where it smoothly drops to zero. The finite cylinder (right)
is effectively capped by repulsive hemispheres (not shown).
}
\label{fig:models_protein}
\end{figure}

\subsection{Protein Models}

\subsubsection{Cylinder of infinite length}
\label{sec:protein_infinite}

The first type of model protein considered here is the infinite straight
cylinder originally introduced by West et al.~\cite{wesbro:09}.  (see
Fig.~\ref{fig:models_protein}, left), {\em i.e.}, the protein cannot tilt.  The
axis of the cylinder is always aligned parallel to the $z$ axis of the
simulation box.  Since this model protein spans the whole simulation box in the
$z$ direction and due to the periodic boundary conditions, the term "infinitely
long cylinder" is justified. The interaction of this simple model proteins and
the lipid or solvent beads has repulsive contributions, which are described by
a radially shifted and truncated Lennard-Jones potential
\begin{equation}
V_{\tn{rep}} (r) = \left\{
\begin{array}{rl}
V_{\tn{LJ}} \left( \frac{r - \sigma_0}{\sigma} \right) -
V_{\tn{LJ}}(1) & 
\tn{if $r - \sigma_0 < 0$}\\
0 & \tn{otherwise}\\
\end{array}  \right . \, ,
\label{equ:V_rep}
\end{equation}
where $r = \sqrt{x^2 + y^2}$ denotes the distance of the interaction
partners in the $xy$ plane, $\sigma$ is given by $\sigma = (\sigma_t
+ \sigma_i)/2$ for interactions with beads of type $i$ ($i = h$, $t$,
and $s$ for head, tail, and solvent beads, respectively), $\sigma_0 =
\sigma_t$, and $V_{\tn{LJ}}$ has been defined above
(Eq.~\eqref{equ:V_LJ}). The direct protein-protein interactions have the
same potential (Eq.~\eqref{equ:V_rep}) with $\sigma = \sigma_t$ and
$\sigma_0 = 2\sigma_t$. In SI-units, the cylinders thus have a diameter
of 2 nm, which roughly corresponds to the diameter of the 8-stranded 
$\beta$-barrel OmpA~\cite{albjoh:02}.

In addition, protein cylinders attract tail beads on a hydrophobic
section of length $L$. This is described by an additional attractive
potential that depends on the $z$ distance between the tail bead and
the protein center. The total potential reads
\begin{equation}
V_{pt}(r, z) = \epsilon_{pt} \left( V_{\tn{rep}}(r) +
V_{\tn{attr}}(r) \times W_P(z) \right)\, ,
\label{equ:V_pt}
\end{equation}
with the attractive Lennard-Jones contribution
\begin{equation}
V_{\tn{attr}}(r) = \left\{
\begin{array}{rl}
V_{\tn{LJ}}(1) - V_{\tn{LJ}}(2) & 
\tn{if $r - \sigma_0 < \sigma$}\\
V_{\tn{LJ}}\left( \frac{ r - \sigma_0}{\sigma} \right) - V_{\tn{LJ}}(2) &
\tn{if $\sigma < r - \sigma_0 < 2 \sigma$}\\
0 & \tn{otherwise}
\end{array} \right . ,
\label{equ:V_attr}
\end{equation}
and a weight function $W_P(z)$, which is unity on a stretch of length
$2 l =  L - 2 \sigma_t$ and crosses smoothly over to zero over a
distance of approximately $\sigma_t$ at both sides. Specifically, we
use
\begin{equation}
W_P = \left \{
\begin{array}{rl}
1 & \tn{if $|z| \leq l$}\\
\cos^2 \left( \frac{3}{2} \left( |z| - l \right) \right) & \tn{if $l < |z| < l +
\frac{\pi}{3}$}\\
0 & \tn{otherwise}
\end{array} \right. .
\label{equ:W_P}
\end{equation}
The hydrophobicity of the protein is tuned by the parameter
$\epsilon_{pt}$. 

\subsubsection{Cylinder of finite length}

The second type of protein is also a straight cylinder, where the hydrophobic
section of length $L$ is modeled analogously to the infinite cylinder. But now
the cylinder is allowed to tilt away from the $z$ direction of the simulation
box (Fig.~\ref{fig:models_protein}, right). The protein has finite length $L$
and is capped at both ends by effectively repulsive (''hydrophilic'') hemispheres. 
More precisely, the protein is parametrized by a line of length $L$, and the
interactions between lipids and proteins are given by Eq.~(\ref{equ:V_pt}) with
$\sigma_0 = \sigma_t$ as before, where $r$ is now the shortest distance between
the center of a lipid bead and the protein line.  The interactions between two
proteins are purely repulsive and given by (\ref{equ:V_rep}) with 
$\sigma_0 = 2\,\sigma_t$ and $r$ the minimum distance between the two 
interacting protein lines.

Even though these cylinders are allowed to tilt, they basically stay
aligned with the membrane normal in our model bilayers. The
distribution of solid angles assumes a maximum at tilt zero, and the
average tilt angles $\langle \alpha \rangle$ for the parameters used
in the present work are less than 15 degrees, as shown in
Table~\ref{tab:tilt}. 
\begin{table}
\begin{tabular}{c|cc}
L [$\sigma_t$] & $\epsilon_{pt}=3\,\epsilon$ & $\epsilon_{pt} = 6\,\epsilon$ \\
\hline 
4 & 14.1 $\pm$ 2.1 & 12.4 $\pm$ 2 \\
6 & 12.5 $\pm$ 0.8 & 11.9 $\pm$ 2.2 \\
8 & 11.1 $\pm$ 1.2 & 7.74 $\pm$ 1.15
\end{tabular}
\caption{Average tilt angle $\langle \alpha \rangle$ of single 
spherocylinders in membranes.}
\label{tab:tilt}
\end{table}
These tilt angles should be regarded as valid in the general 
case, since we are not aiming at modeling any specific protein. 
Our results are compatible with an earlier simulation study by Venturoli 
\emph{et al.}\cite{vensmi:05}, where the average tilt angle of proteins with both 
negative and positive mismatch of up to $40\%$ was also only slightly 
affected by the mismatch. Only when a considerable positive mismatch 
of more than $70\%$ was present, a significant increase in average 
tilt angle for model proteins of small aspect ratio (diameter to
length) was observed.

Since a protein is "rough"\ on an atomic scale, representing a complex
structure like a protein as a simple smooth cylindrical object may seem to be a
rather crude approach. Nevertheless, our way of modeling the proteins can be
justified by the fact that e.g. $\alpha$-helices are packed with vanishing
little free space within the helices \cite{boh:09} and are therefore fairly
smooth on the scale of $\sim 10\,\tn{\AA}$. There are no large cavities into
which chains or even whole molecules would fit \cite{scocoe:83}.

\subsection{Simulation Method}

We have carried out Monte Carlo simulations at constant pressure $P = 2.0 \,
\epsilon/\sigma_t^3$ and constant temperature $T$ with periodic boundary
conditions in a simulation box of variable shape and size. Thus, we are
performing Monte-Carlo simulations in an $N P T$ ensemble with effective
Hamiltonian \begin{equation} \label{equ:hamiltonian_gamma} H_{\mathrm{eff}} = U
+ P V - N k_B T \ln(V/V_0) \quad , \end{equation} where $U$ is the interaction
energy, $V$ the volume of the simulation box, $V_0$ an arbitrary reference
volume and $N$ the total number of beads.
 
In practice, three main types of Monte-Carlo moves were proposed and then
accepted or rejected according to a Metropolis criterion, namely 1)
translational local moves of the lipid beads, 2) protein translation and
rotation moves, and 3) global moves which change the volume of the simulation
box or its shape.  Most of these moves have been discussed earlier
\cite{schdue:07}.  Here we only sketch the algorithm for the new protein
rotation moves.  The new trial direction $\widehat{\ve{d}}'_p$ of the protein
is generated in several steps.  First a vector $\widehat{\ve{u}}$, which lies
randomly distributed on a unit sphere, is generated. Choosing a random vector
on the surface of a unit sphere is efficiently done by applying an algorithm
proposed by Marsaglia \cite{mar:72}: Two random variables $r_1$ and $r_2$
within the interval $(-1,1)$ are generated and $\zeta^2 = r_1^2 + r_2^2$ is
calculated. If $\zeta^2 > 1$ the random numbers are discarded and a new pair
$r_1$, $r_2$ has to be generated. For $\zeta^2 < 1$ the components of the
random unit vector in Cartesian coordinates are given by
\begin{equation}
\begin{split}
\widehat{u}_x &= 2 r_1 \sqrt{1 - \zeta^2}\\
\widehat{u}_y &= 2 r_2 \sqrt{1 - \zeta^2}\\
\widehat{u}_z &= 1 - 2 \zeta^2 \, .
\end{split}
\label{equ:marsaglia}
\end{equation}  
If the new direction of the protein was chosen to be $\widehat{\ve{u}}$, the
acceptance rate of the move would be prohibitively small. Therefore,
$\widehat{\ve{u}}$ is scaled down by $\Delta_{\tn{tilt}}$ and the vector
$\ve{t} = \Delta_{\tn{tilt}} \widehat{\ve{u}}$ is added to the tip of the
normalized direction $\widehat{\ve{d}}_p$ of the protein.  Finally, normalizing
the resulting vector $\ve{d}'_p = \widehat{\ve{d}}_p + \ve{t}$ to
$\widehat{\ve{d}'}_p$ gives the trial direction of the protein used to
calculate the energy change for the Metropolis criterion. By varying the
scaling factor $\Delta_{\tn{tilt}}$ during the "prerun"\ the acceptance rate of
the tilt moves can be adjusted to the desired rate.  Effectively, this means
adjusting the opening angle of the cone around $\widehat{\ve{d}}_p$, which
limits the maximum tilt angle $\theta_{p,\tn{max}}$ of the move.  

During one Monte-Carlo step ($\MCS$) there is on average one attempt to move
each bead and protein, and one attempt to rotate the protein. Since the global
moves require rescaling of all particle coordinates, which is rather expensive
from a computational point of view, they are performed only every 50th $\MCS$
on average. 

\subsection{Elastic Theory}

\label{sec:theory}

In the following, we briefly sketch the elastic theory with which we compare
the simulation results \cite{araber:96,brabro:06,wesbro:09}, and which has
proved to describe very well the properties of our model membranes in the 
fluid state \cite{wesbro:09}. The membrane is treated as a system of coupled 
monolayers, which fluctuate subject to the constraint that
the volume of lipids is locally conserved. It is also considered to be
almost flat, {\em i.e.}, the positions of both monolayers can be parametrized 
by single-valued functions $h_i(x,y)$. Here we are mainly interested in the
thickness deformations, which we characterize by the locally smoothed 
deformation profiles $\Phi_{\tn{el}}(x,y)$ of monolayers about the unperturbed
value $t_0$ ( {\em i.e.}, the total monolayer thickness is $t_0 + \Phi_{\tn{el}}$).
The free energy of monolayer thickness deformations is then written as 
\cite{brabro:07}
\begin{equation}
\begin{split}
F = \int {\rm d}^2r \Bigg\{
\frac{k_A}{2 t_0^2} \Phi_{\tn{el}}^2 
+ 2 k_c \left( c_0  + \zeta \frac{\Phi_{\tn{el}}}{t_0} \right) \Delta \Phi_{\tn{el}}\\
+ \frac{k_c}{2} (\Delta \Phi_{\tn{el}})^2 
+ k_G \det(\partial_{ij} \Phi_{\tn{el}}) \Bigg\}\,,
\end{split}
\label{equ:elastic}
\end{equation}
where $k_c$ and $k_A$ are the bending and compressibility modulus of the
bilayer, $c_0$ is the spontaneous curvature, $\zeta$ a parameter related to the
spontaneous curvature, and $k_G$ the Gaussian curvature. All these
parameters have been determined for our model membranes in earlier work from
analyses of the membrane fluctuations and stress profiles \cite{wesbro:09}. 
We note that the terms $c_0$ and $k_G$ can be rewritten as pure surface terms.
Protein inclusions distort the membranes by imposing a fixed thickness 
at their boundaries \cite{araber:96}, $\Phi_{\tn{el}} = t_R$, and may introduce 
additional surface fields, which enter in the same way as the spontaneous
curvature term \cite{wesbro:09} and thus effectively renormalize $c_0$.
Hence the effect of proteins on the membrane can be characterized by the 
two parameters $t_R$ and $\tilde{c}_0$, the renormalized curvature. 
They have been determined for our model in earlier work \cite{wesbro:09}
from the distortion profiles of single membranes about infinitely long 
straight cylinders. Table~\ref{tab:parameters} summarizes the values
of all elastic parameters for the simulation models studied in this
work.

The elastic free energy, Eq.~(\ref{equ:elastic}), can be minimized analytically
for membranes containing only a single cylindrical protein. For two proteins,
it must be minimized numerically. To this end, the free energy integral was
discretized in real space using a square grid of spatial discretization $h =
0.25\,\sigma_t$, a system size of $50 \times 50\,\sigma_t^2$ and a second-order
difference scheme to evaluate the derivatives. The boundary condition was
implemented by setting $\phi = t_R$ inside the inclusion. The minimization was
done via a steepest descent method, using a relaxation scheme \cite{schsch:95}.
This procedure gave deformation profiles with which we could compare the
simulation results (see below).

\begin{table}
\begin{tabular}{|c|c|c|}
\hline 
Parameter & $L$ & Value \\
\hline 
$k_c$ &  all & 6.2 $\epsilon$ \\
$\zeta/t_0$ & all &0.15 $\sigma_t^{-2}$ \\
$k_A/t_0^2$ & all & 1.3 $\epsilon/\sigma_t^4$ \\
$k_G$ & all & -0.26 $\epsilon$ \\
\hline 
$t_R$ & 4 $\sigma_t$ & -0.94 $\sigma_t$ \\
$t_R$ & 6 $\sigma_t$ & 0.3 $\sigma_t$ \\
$t_R$ & 8 $\sigma_t$ & 1.44 $\sigma_t$ \\
\hline 
$\tilde{c}_0$ & 4 $\sigma_t$ & -0.11 $\sigma_t^{-1}$ \\
$\tilde{c}_0$ & 6 $\sigma_t$ & 0.05 $\sigma_t^{-1}$ \\
$\tilde{c}_0$ & 8 $\sigma_t$ & 0.22 $\sigma_t^{-1}$ \\
\hline
\end{tabular}
\caption{Elastic parameters used in the theoretical calculations.
Taken from Ref.~\protect\cite{wesbro:09}.}
\label{tab:parameters}
\end{table}

\section{Simulation Results} 
\subsection{Lipid-mediated interactions between inclusions}
\label{sec:pmf}

We now turn to discussing our simulation results for the membrane-induced
interactions between cylindrical inclusions in the bilayer. 
In this subsection, we focus on the effect of protein orientation
fluctuations on the effective interactions between the model proteins, 
{\em i.e.}, the PMF.

The radial distribution function $g(r)$ as a function of the protein-protein
distance $r$ was obtained from simulation runs using the technique of
successive umbrella sampling \cite{virmue:04} combined with a reweighting
procedure. As starting configurations we used equilibrated systems with 750 to
760 lipids and two simple transmembrane proteins of diameter $3\,\sigma_t$. A
first estimate of $g(r)$  was obtained during pre-runs of length
$2\times10^6\,\MCS$.  Then, biased runs of $3\times10^6\,\MCS$ were performed
to improve the statistics of configurationally less frequent protein-protein
distances. In the case of tiltable proteins, $10 \times 10^6\,\MCS$ had to be
performed to 
achieve comparable accuracy in the histogram data.

After removing the bias from these results and combining the overlapping
distributions the effective potential $w(r) = -k_B T \ln g(r)$ can be
extracted.  At small inclusion-inclusion distances direct interactions of
proteins  and, in the case of long proteins, additional depletion induced
attraction due to the solvent particles come into play. Since our main interest
lies in the study of lipid-mediated medium and long ranged interactions, these
parts of the curves have been cut off. Our sampling procedure resulted in a
statistical error of about $\pm 0.3 \epsilon$ within each umbrella window. The
distance range between $4$ and $10\,\sigma_t$ was typically covered by 7-8 windows
(with larger windows at larger distances), which gives an accumulated error of
$\pm 2\,\epsilon$ in the first minimum of the PMFs in the worst case, and of
at most $\pm 1\,\epsilon$ at distance $6\,\sigma$.

The hydrophobicity of the protein, {\em i.e.}, the interaction strength between
the hydrophobic core of the membrane and the hydrophobic part of the inclusion,
is tuned by the parameter $\epsilon_{pt}$.  Only for sufficiently high values
of $\epsilon_{pt}$ do the model proteins locally distort the bilayers.  In the
following we will consider two cases, weakly hydrophobic proteins with
$\epsilon_{pt} = 3.0\,\epsilon$ and strongly hydrophobic proteins with
$\epsilon_{pt} = 6.0\,\epsilon$. In the first case lipid-protein contact is
approximately energetically equivalent to lipid-lipid contact, whereas in the
second case lipid-protein contact is highly preferred. Weakly hydrophobic proteins 
only induce small perturbations of the lipid environment, strongly hydrophobic
proteins lead to strong deformations. 

Furthermore, the hydrophobic length $L$ has been varied from $L=4\,\sigma_t$ 
to $L=8\,\sigma_t$. The hydrophobic thickness of the membrane is $6\,\sigma_t$,
hence $L=4\,\sigma_t$ corresponds to a negatively mismatched protein, and
$L= 8\,\sigma_t$ to a positively mismatched protein.

\begin{figure}
\begin{center}
\includegraphics[angle=0,width=\columnwidth]{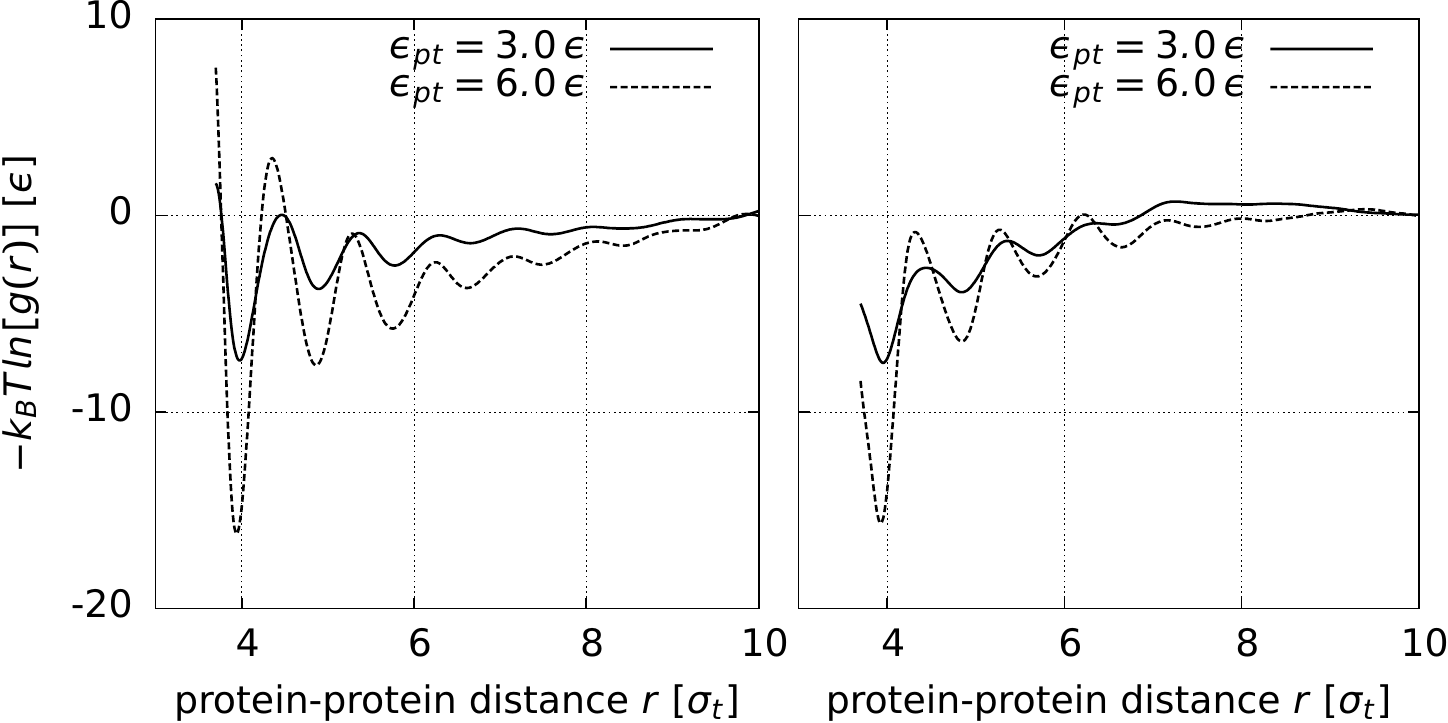}
\caption{
Potential of mean force of two isolated proteins as a function of the
protein distance at medium (solid) and strong (dashed)
hydrophobicity $\epsilon_{pt}$ for proteins with negative hydrophobic
mismatch ($L=4\,\sigma_t$). The proteins of the left graph were always 
aligned parallel to the $z$ axis of the simulation box, whereas proteins 
in the right graph were allowed to tilt.}
\label{fig:T1p3_pl2p0_smt4p0_ept3-6_compare}
\end{center}
\end{figure}

\begin{figure}
\begin{center}
\includegraphics[angle=0,width=\columnwidth]{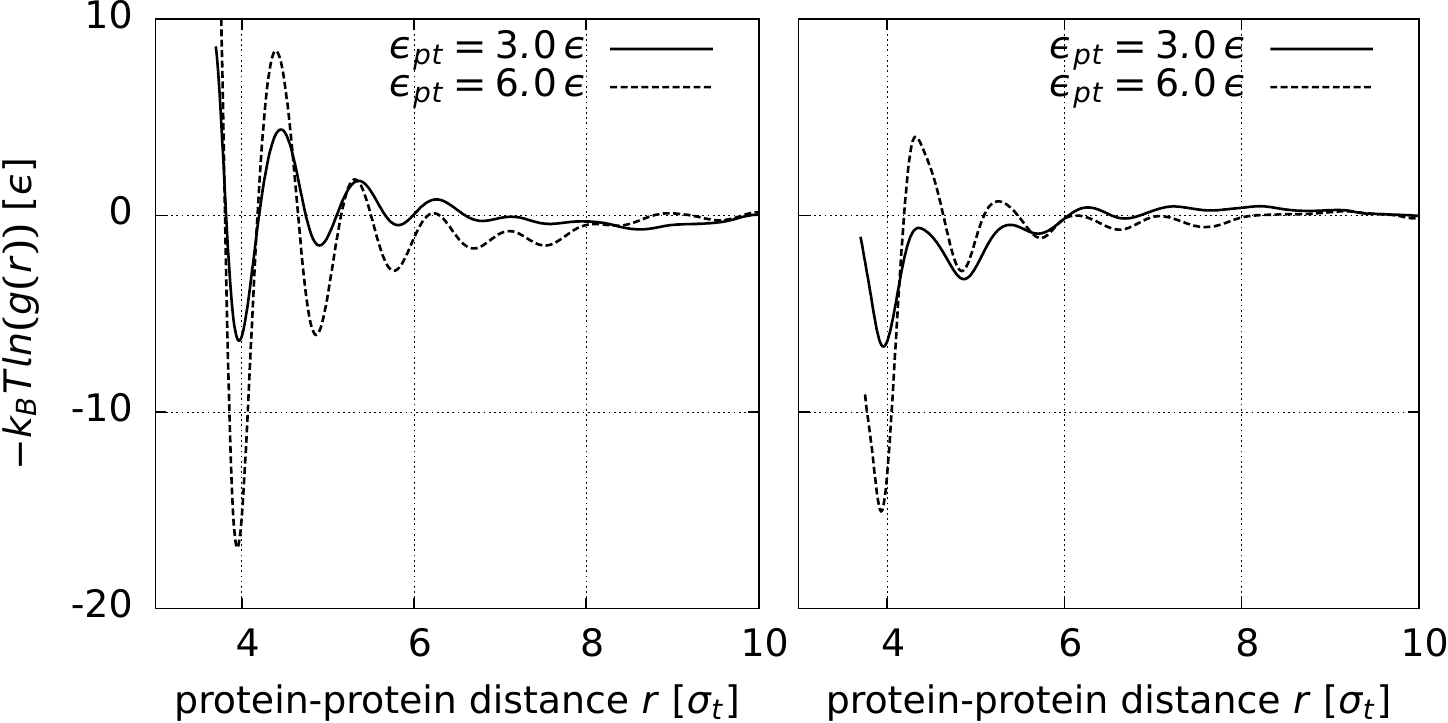}
\caption{
Potential of mean force of two isolated proteins as a function of the
protein distance at medium (solid) and strong (dashed)
hydrophobicity $\epsilon_{pt}$ for hydrophobically matching proteins
($L = 6 \sigma_t$).
(left: aligned protein, right: tiltable proteins).}
\label{fig:T1p3_pl4p0_smt6p0_ept3-6_compare}
\end{center}
\end{figure}

\begin{figure}
\begin{center}
\includegraphics[angle=0,width=\columnwidth]{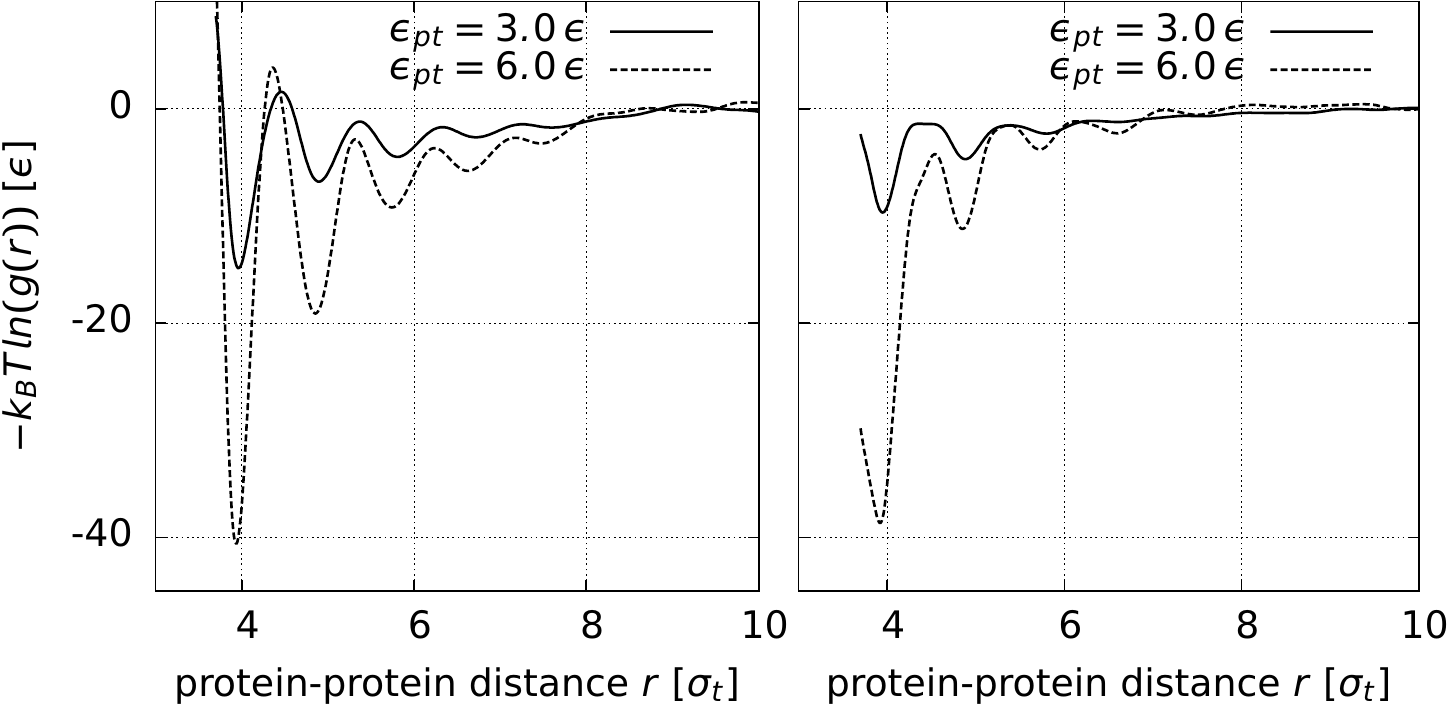}
\caption{
Potential of mean force of two isolated proteins as a function of the
protein distance at medium (solid) and strong (dashed)
hydrophobicity $\epsilon_{pt}$ for proteins with positive hydrophobic
mismatch ($L=8\,\sigma_t$) (left: aligned protein, right: tiltable proteins).}
\label{fig:T1p3_pl6p0_smt8p0_ept3-6_compare}
\end{center}
\end{figure}

The simulation results are compiled in Figs.
\ref{fig:T1p3_pl2p0_smt4p0_ept3-6_compare},
\ref{fig:T1p3_pl4p0_smt6p0_ept3-6_compare}, and
\ref{fig:T1p3_pl6p0_smt8p0_ept3-6_compare}.  The left and right graphs in each
figure correspond to straight and tiltable cylinders, respectively. One
immediately notices the first effect of orientational fluctuations: Due to
lipid packing in the vicinity of the inclusions the curves feature oscillations
with a wavelength of approximately $1\,\sigma_t$, i.e. the diameter of the
lipid tail beads. The higher $\epsilon_{pt}$ the more pronounced the lipid
layering. This holds both for straight and tiltable proteins. However, the
oscillations are significantly reduced in the latter case. Hence orientation
fluctuations reduce the effect of lipid layering on the protein-protein
interactions.

Next we discuss separately the PMFs for strongly and weakly hydrophobic
proteins.

\subsubsection{Strongly hydrophobic proteins}


A general feature that straight and tiltable protein have in
common at $\epsilon_{pt} = 6.0\,\epsilon$  is an enhanced tendency to aggregate
for hydrophobically mismatched proteins, both for positive or negative mismatch
(dashed curves in Figs.~\ref{fig:T1p3_pl2p0_smt4p0_ept3-6_compare} and
\ref{fig:T1p3_pl6p0_smt8p0_ept3-6_compare}). The effective interactions between
mismatched proteins have an attractive long-range contribution which is not
detectable for hydrophobically matching proteins
(Fig.~\ref{fig:T1p3_pl4p0_smt6p0_ept3-6_compare}). Compared to straight
proteins, however, the range of this additional attraction is reduced for
proteins with orientational fluctuations, and the attraction essentially
vanishes within the error at distances beyond $r \sim 7\,\sigma_t$.

The first and deepest attractive minimum of the PMFs slightly below $r =
4\,\sigma_t$ lies at approximately the same energy for both types of models,
i.e. between $-17$ and $-15\,\epsilon$ for proteins of length $L = 4\,\sigma_t$
and $6\,\sigma_t$ and about $-40\,\epsilon$ for proteins of length $L =
8\,\sigma_t$. This minimum is related to lipid bridging and a consequence of
the strong attraction between lipids and proteins.  At distance $r
\sim 4\,\sigma_t$, one layer of lipids is in contact with both proteins, thus
stabilizing the conformation at that distance. The lipid bridging energy
increases with increasing hydrophobic contact area between lipid tails and
proteins. Compared to hydrophobically matching proteins, it is thus reduced for
negatively mismatched proteins and enhanced for positively mismatched proteins.
In addition, both negatively and positively mismatched proteins have the
above-mentioned attractive interactions due to the elastic deformation of the
layer. Compared to hydrophobically matching proteins, the contributions of
lipid bridging and hydrophobic mismatch are opposite for negatively mismatched 
proteins, and cumulative for positively mismatched proteins. 
As a result the first minimum has approximately the same energy for
proteins with negative hydrophobic mismatch and matching proteins, whereas
proteins with positive mismatch show the strongest attraction.

\subsubsection{Medium hydrophobicity}

For weakly hydrophobic proteins with $\epsilon_{pt}= 3.0\,\epsilon$, the effect
of hydrophobic mismatch at short distances is much less pronounced than in the
case of strongly hydrophobic proteins. This is especially true for proteins
with fluctuating orientations. In the latter case, the first minimum has almost
the same depth for all types of mismatch, and the PMFs for different $L$ are
not further apart than $3\,\epsilon$ (solid curves in the right graphs of
Fig.~\ref{fig:T1p3_pl2p0_smt4p0_ept3-6_compare},
\ref{fig:T1p3_pl4p0_smt6p0_ept3-6_compare}, and
\ref{fig:T1p3_pl6p0_smt8p0_ept3-6_compare}). 

At intermediate distances, however, the interaction between proteins with
fluctuating orientations is found to depend crucially on the type of mismatch:
For positively mismatched proteins, it has a long range attractive contribution
similar to that observed for strongly hydrophobic proteins.  For negatively
mismatched proteins, this contribution is still present at distances $r <
7\,\sigma_t$, but then it turns around and gives way to a {\em repulsive}
interaction. The repulsive barrier is small, but significant within the error
of $\pm 0.5\,\epsilon$ at distance $8\,\sigma_t$. The behavior for matching proteins is
intermediate -- the layering interactions are superimposed by an attractive
contribution up to $r \sim 6\,\sigma_t$, which disappears within the error at
larger distances.  We note that this behavior is in marked contrast both to the
behavior of strongly hydrophobic proteins and of weakly hydrophobic, but
straight inclusions. In the latter cases, one observes a purely attractive
long-range hydrophobic mismatch interaction, which vanishes for hydrophobically
matched proteins.

In fact, the elastic theory does predict a repulsive barrier in the interaction
free energy of two transmembrane proteins for positive and negative hydrophobic
mismatch \cite{wesbro:09}. The maximum of this small repulsive barrier should
be found in the range of $r \sim 6\,\sigma_t$ to $8\,\sigma_t$ and can be
associated with a peak, {\em i.e.}, a soft mode, in the spectrum of thickness
fluctuations of the pure bilayer.  One may speculate that the repulsive
shoulders observed here are caused by the same effect. However, our
simulations results show no repulsive barrier for proteins of length $L=
8.0\,\sigma_t$. The attractive range for positively mismatched proteins rather
extends up to a protein-protein separation of about $9\,\sigma_t$ and levels
off without any observable repulsive region. Furthermore, the attractive
behavior of hydrophobically matching proteins is not predicted by the elastic
theory either. 


Other comparable simulations on tiltable proteins have been performed by
Schmidt et al. Both for negative and positive hydrophobic mismatch they observe
a long-range, lipid-mediated attraction. In the case of zero mismatch this
interaction is less attractive \cite{schgui:09}.  Further, their PMFs exhibit
an oscillating "fine structure", which they attribute to the discreteness of
the membrane.

In another study, de Meyer et al. have compared PMFs for proteins which were
allowed to tilt, and proteins which were always aligned parallel to the
$z$-axis \cite{demven:08}. Their results for proteins comparable to our capped
cylindrical inclusions of diameter $D_p = 3\,\sigma_t$ can be summarized as
follows: In the case of positive mismatch the long-range interaction between
two proteins should be influenced by the degree of protein tilt. If the protein
was not allowed to tilt, they found an attractive region for small distances,
followed by a small repulsive barrier and a shallow and broad minimum at large
protein separation.  If tilting of proteins was allowed, only a
small, but broad repulsive barrier at medium inter-protein distances was found
after the typical attractive range at small distances for both positive and
negative mismatch. The shallow attractive region at larger distance has
vanished. In our model of long cylindrical proteins without tilt a shallow
repulsive barrier at medium distances was also present, but only at very weak
hydrophobicity of $\epsilon_{pt} = 1.0\,\epsilon$ \cite{wesbro:09}. Other effects
due to lipid layering were much less pronounced in the study of de~Meyer et
al., since they were using soft, merely repulsive potentials. But at very small
protein separations they did observe slight oscillations in the PMFs, which
they assigned to the free energy needed to remove lipids in between the
proteins. 

\subsection{Profiles around two proteins}
\label{sec:def_prof2D}

After having discussed the effective lipid-mediated interactions between two
transmembrane inclusions, we now turn to the closer investigation of structural
lipid rearrangements around two model proteins at small and medium separation. 
In this section, we will consider strongly hydrophobic proteins with
$\epsilon_{pt} = 6\,\epsilon$.

As we have seen in the previous section, one characteristics of our lipid model
with its hard-core, Lennard-Jones type interactions is the presence of lipid
packing phenomena, which are usually much less prominent in systems modeled
with soft dissipative particle dynamics potentials \cite{wesbro:09}. Especially
in the area surrounding proteins one can expect ordering of the
lipids, which is rather distinct from the situation further away from
inclusions in the lipid bulk, and which may affect the interaction with other
inclusions.  Here, two-dimensional profiles give useful insight into the structure
of membrane-characterizing quantities. To determine them, we have simulated
equilibrated bilayers containing two capped proteins, keeping the positions
of the proteins fixed, but allowing for tilt fluctuations.
As before, we compare systems with cylindrical proteins of varying
hydrophobicity, hydrophobic mismatch and protein-protein distance.

The two-dimensional profiles can be separated into two regions: In the inner
region between the two inclusions, the profiles reflect the combined
effect of both proteins. In the outer region, the influence of one protein
is geometrically screened, and the profiles mainly reproduce the perturbations
induced by one single protein.

\subsubsection{Areal tail bead density} \label{sec:areal_tail_density}

\begin{figure}[] 
\begin{center} 
\begin{tabular}{ c c c } 
$d = 3.99\,\sigma_t$& \hspace{0.05\columnwidth}  & $d = 6.45\,\sigma_t$ \\
\includegraphics[width=0.45\columnwidth]{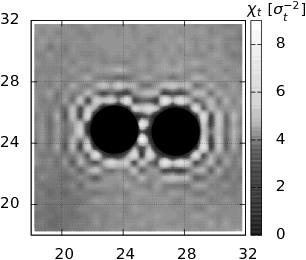} &
\hspace{0.05\columnwidth} &
\includegraphics[width=0.45\columnwidth]{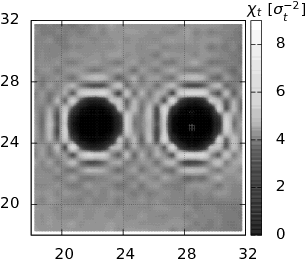}\\
\end{tabular} 
\end{center}
\caption{Areal tail density profiles for systems containing
two translationally immobile hydrophobically matched proteins ($L=6\,\sigma_t$) 
for two different protein distances, obtained by binning within a quadratic
grid with bin size $0.5 \times 0.5\,\sigma_t^2$. The profiles reflect the
underlying grid.} 
\label{fig:prof2D_areal_tail_density_smt6p0}
\end{figure}

First, the areal density of tail beads $\chi_t$ around two proteins is
investigated. In the following graphs, the values are always given with
respect to one monolayer, i.e. the average number of tail beads
obtained for each bin lying in the $xy$ plane was divided by $2$. The 
reference value of $\chi_t$ in the unperturbed bilayer is simply the
number of tail beads per lipid divided by the average area per lipid 
at $T = 1.3\,k_B T$, i.e. $\chi_t^0 = 6 / 1.38\,\sigma_t^2 =
4.3\,\sigma_t^{-2}$.

Both length and height of a single bin is set to $0.5\,\sigma_t$.
Therefore each sampled cell has an area of $0.25\,\sigma_t^2$
only. At this resolution interesting details of the lipid structure in
the vicinity of the inclusions are revealed.

Not surprisingly, negative hydrophobic mismatch, which goes along with thinning
of the surrounding bilayer, leads to a reduction of the areal tail density,
whereas positive hydrophobic mismatch enhances the areal tail density along with
the membrane thickness. The fine structure exhibits additional features. In
the following, we will focus on hydrophobically matching proteins of length $L
= 6\,\sigma_t$.  The layering of the lipids around the proteins, which gave
rise to the oscillatory behavior of the PMFs in the previous section, manifests
itself clearly in concentric rings with enhanced (light) and reduced (dark)
areal tail bead density (Fig.~\ref{fig:prof2D_areal_tail_density_smt6p0}). The
variations in areal density are usually in the order of $20\%$ compared to the
average value at larger distance from the inclusions. If the proteins are
close to each other (Fig.~\ref{fig:prof2D_areal_tail_density_smt6p0}, left),
they pin the lipids entirely {\em i.e.}, the areal tail density exhibits 
sharp local peaks at two well-defined positions between the two proteins.

\subsubsection{Thickness profiles}

\label{sec:thickness_profiles}
\begin{figure}[] 
\centering
\begin{tabular}{c c c }
	& simulation	& theory \\
\rotatebox[origin=l]{90}{$\quad \quad \quad r = 3.96\,\sigma_t$} &
\includegraphics[width=0.45\columnwidth]{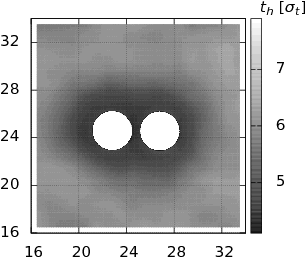} &
\includegraphics[width=0.45\columnwidth]{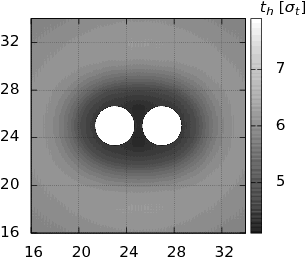}\\
\rotatebox[origin=l]{90}{$\quad \quad \quad r = 6.56\,\sigma_t$} &
\includegraphics[width=0.45\columnwidth]{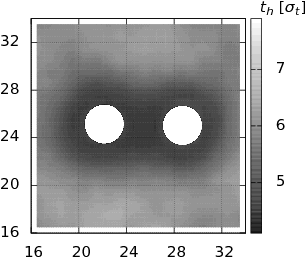} &
\includegraphics[width=0.45\columnwidth]{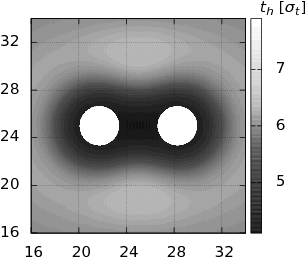}\\
\rotatebox[origin=l]{90}{$\quad \quad \quad r = 8.16\,\sigma_t$} &
\includegraphics[width=0.45\columnwidth]{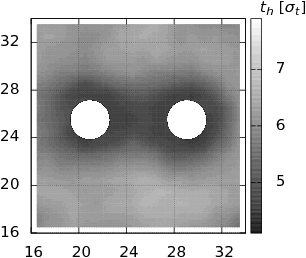} &
\includegraphics[width=0.45\columnwidth]{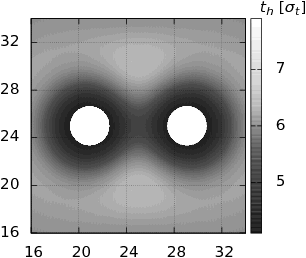}\\
\end{tabular}
\caption{Thickness profiles around two translationally immobile
cylindrical proteins with negative hydrophobic mismatch ($L = 4.0\,\sigma_t$).
Left: Simulations, Right: Elastic theory}
\label{fig:prof2D_thickness_smt4p0}
\end{figure}

\begin{figure}[] 
\centering
\begin{tabular}{c c c }
	& simulation	& theory \\
\rotatebox[origin=l]{90}{$\quad \quad \quad r = 3.88\,\sigma_t$} &
\includegraphics[width=0.45\columnwidth]{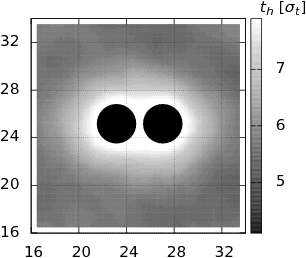} &
\includegraphics[width=0.45\columnwidth]{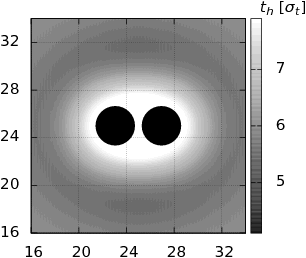}\\
\rotatebox[origin=l]{90}{$\quad \quad \quad r = 6.57\,\sigma_t$} &
\includegraphics[width=0.45\columnwidth]{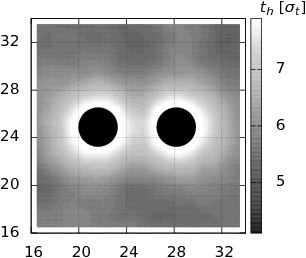} &
\includegraphics[width=0.45\columnwidth]{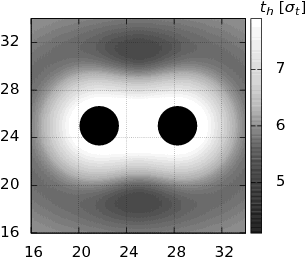}\\
\rotatebox[origin=l]{90}{$\quad \quad \quad r = 8.97\,\sigma_t$} &
\includegraphics[width=0.45\columnwidth]{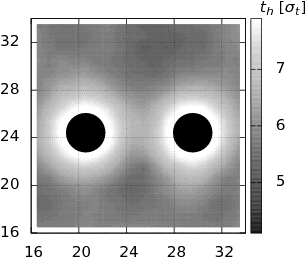} &
\includegraphics[width=0.45\columnwidth]{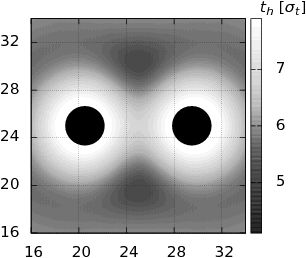}\\
\end{tabular}
\caption{Thickness profiles around two translationally immobile
cylindrical proteins with positive hydrophobic mismatch ($L = 8.0\,\sigma_t$).
Left: Simulations, Right: Elastic theory
}
\label{fig:prof2D_thickness_smt8p0}
\end{figure}

\begin{figure}[] 
\centering
\begin{tabular}{c c c }
	& simulation	& theory \\
\rotatebox[origin=l]{90}{$\quad \quad \quad r = 3.99\,\sigma_t$} &
\includegraphics[width=0.45\columnwidth]{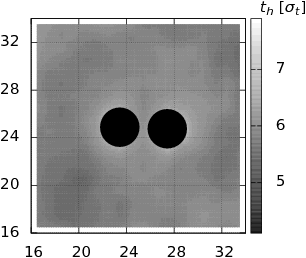} &
\includegraphics[width=0.45\columnwidth]{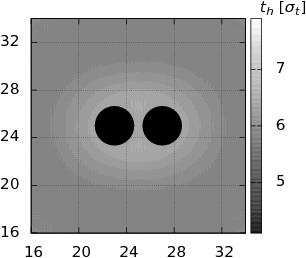}\\
\rotatebox[origin=l]{90}{$\quad \quad \quad r = 6.45\,\sigma_t$} &
\includegraphics[width=0.45\columnwidth]{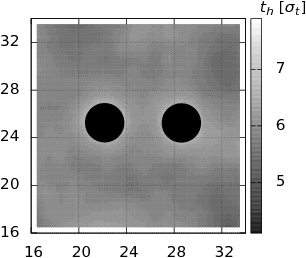} &
\includegraphics[width=0.45\columnwidth]{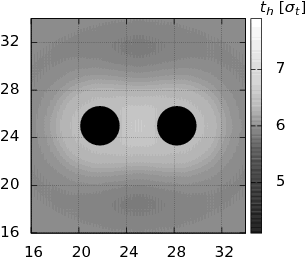}\\
\end{tabular}
\caption{Thickness profiles around two translationally immobile
cylindrical proteins without mismatch ($L = 6.0\,\sigma_t$).
Left: Simulations, Right: Elastic theory.
}
\label{fig:prof2D_thickness_smt6p0}
\end{figure}

Next, we inspect the thickness profiles around two cylindrical inclusions at
fixed distances.  Specifically, we consider short distances, slightly below a
protein-protein separation of $\sim 4\,\sigma_t$, medium distances at a
separation of $\sim 6.5\,\sigma_t$, and larger separations lying between $8$
and $9\,\sigma_t$.  A regular grid of cells with an area of $\Delta x \Delta y
= 1.0\,\sigma_t^2$ was used to subdivide the system.  Averaging of the membrane
thickness was done within each quadratic bin
thus obtained. If necessary, the systems were rotated such that the connecting
line between the centers of the proteins was parallel to the $x$-axis for
better comparison and visualization purposes. Light shading indicates thick
membranes, dark shading thin membranes. For comparison, we have also calculated
the same profiles with the elastic model described earlier
(Sec.~\ref{sec:theory}), using the elastic parameters obtained in
Ref.~\cite{wesbro:09} (Table~\ref{tab:parameters}). The results are shown in
Figs.~\ref{fig:prof2D_thickness_smt4p0}, \ref{fig:prof2D_thickness_smt8p0}, and
\ref{fig:prof2D_thickness_smt6p0}.

Both for small and medium protein-protein separations the perturbations induced
by the proteins overlap.  This amplifies the reduction or increase of thickness
which would be caused by a single protein of negative or positive hydrophobic
mismatch, respectively. At large distances, the proteins are surrounded by
separate perturbation shells.

Fig.~\ref{fig:prof2D_thickness_smt4p0} shows the results for proteins with
negative mismatch. For small and medium protein-protein distances the bilayer
thickness is smallest in the region between the proteins
(Fig.~\ref{fig:prof2D_thickness_smt4p0}, top left and middle left), which is in
remarkable agreement with the predictions of the elastic theory
(Figs.~\ref{fig:prof2D_thickness_smt4p0}, top right and middle right).  For
protein-protein distances $5-7 \sigma_t$, the elastic theory even predicts a
certain amount of "undershooting", {\em i.e.}, the membrane thickness on the
axis between the two proteins is predicted to be smaller than at the protein
surface. This is also found in simulations, albeit to a much lesser extent.

Proteins with positive hydrophobic mismatch ($L = 8.0\,\sigma_t$) are
surrounded by a clearly thickened bilayer (Fig.~\ref{fig:prof2D_thickness_smt8p0}).  
As for negatively mismatched proteins, the membrane perturbations
are enhanced in the region between the proteins. For protein distances
$5-7\,\sigma_t$, the elastic theory predicts "overshooting", {\em i.e.}, 
the thickness is expected to be largest at the point right between the
two proteins. However, the simulation data do not show this effect.
In this respect, positively mismatched proteins show a different behavior
than negatively mismatched proteins.

Finally, the bilayer thickness between hydrophobically matching proteins ($L =
6.0\,\sigma_t$) remains practically unaltered as expected, except in the very
close vicinity of the proteins, where a small annulus of enhanced thickness
develops (Fig.~\ref{fig:prof2D_thickness_smt6p0}, left).  
Elastic theory predicts a slightly enhanced thickness in the 
vicinity of the proteins (Fig.~\ref{fig:prof2D_thickness_smt6p0}, right).

\section{Summary and Discussion}

In summary, we have studied protein-protein interactions in model lipid
membranes, using two types of model proteins: Infinitely long cylinders with
fixed orientation, and finite spherocylinders with fluctuating orientation. 
In general, the interactions are dominated by packing effects and smooth
attractive hydrophobic mismatch interactions. Besides the geometric mismatch
the hydrophobicity of the proteins is a key quantity.

The comparison of the PMFs for the two protein types
allowed to assess the effect of orientation fluctuations. It can be summarized
as follows: The main and most prominent effect of orientation fluctuations is
to reduce the lipid packing effects on the PMFs, {\em i.e.}, the corresponding
short range oscillatory behavior. Furthermore, they slightly reduce the range 
of the hydrophobic mismatch interactions. 
For weakly hydrophobic proteins,
they may also affect qualitatively the shape of the potentials: In the case
of negatively mismatched proteins, orientation fluctuations introduce a
weak repulsive maximum in the PMFs which is not present for orientationally
fixed proteins. This is consistent with earlier results by
de Meyer et al.~\cite{demven:08}, using a different protein model with 
hydrophobic interactions that would be considered 'weak' in our context.

The direct experimental measure of interaction potentials between membrane
proteins is still a challenging task due to the required temporal and spatial
resolution. The "protein" diameter in our study roughly corresponds to that
of gramicidin, for which the effect of hydrophobic mismatch interactions was
verified experimentally in 1999 by Harroun and coworkers \cite{harhel:99:1}. The
hydrophobic mismatch effect has also been investigated by systematic studies of
synthetic model peptides \cite{shabar:02,depkil:03}. In these studies, the
evidence for mismatch-induced interactions was mainly based on the observation
of protein clustering.  Recently, spatially resolved interaction energies
between proteins of positive hydrophobic mismatch have been calculated for
mobile ATP-synthase c-rings in membranes basically consisting of phosphatidyl
glycerophosphate from the probability distribution of the center-to-center
distance by Casuso {\em et al.} \cite{cassen:10}.  After a soft-core short
range repulsion attributed to the structural perturbation of the lipid/protein
organization, a minimum could be observed in the interaction energy. The
authors consider this attractive potential to be caused by membrane-mediated
interactions due to the deformations of the lipid bilayer. The ATP-synthase
c-rings of diameter $65\,\tn{\AA} \pm 5\,\tn{\AA}$ would correspond to proteins
of diameter $10\,\sigma_t$ in our systems. Therefore, direct comparison with
our simulation data is not possible. Nevertheless, the experimental finding of
protein-protein attraction by mismatch-induced deformation is in accordance
with the resulting smooth attractive interaction in our simulations. 

For the case of strong interactions, we have also studied the distribution of
lipid tails in systems containing two proteins and found that two proteins can
form complexes with surrounding lipids, pinning these to their surface. 
We speculate that such effects may also occur in biological environments.
Furthermore, we have compared the thickness profiles in systems containing two proteins with the
elastic theory of two coupled monolayers. We found remarkable agreement between
the theoretical prediction and the simulation results, with only slight
qualitative deviations at intermediate protein distances $\sim 7\,\sigma_t$:
here, the elastic theory predicts "over"- or "undershooting" effects which are
much weaker or not discernible at all in the simulations.  The theoretical
prediction for the PMFs has not been shown here (see \cite{wesbro:09}), because
it is difficult to compare with the simulation data due to the lipid packing
effects. Regarding the smooth hydrophobic mismatch part, the theory is in rough
qualitative agreement with the simulations, except that its predicts a
repulsive barrier at distances $\sim 8\,\sigma_t$ which is not observed in the
simulations. The discrepancy is possibly related to the slight deviations
between theory and simulations in the thickness profiles mentioned above.

\section*{Acknowledgement}

Provision of computing resources by the HLRS (Stuttgart), NIC
(J\"ulich), and PC2 (Paderborn) is gratefully acknowledged.
The configurational snapshots were visualized using VMD \cite{humdal:96}. 
This work was funded in part by the DFG within the 
Sonderforschungsbereich SFB 613, SFB 625, 
and SFB TR 6.

\vspace*{-0.5cm}

\bibliography{jabref_promotion}

\providecommand*{\mcitethebibliography}{\thebibliography}
\csname @ifundefined\endcsname{endmcitethebibliography}
{\let\endmcitethebibliography\endthebibliography}{}
\begin{mcitethebibliography}{42}
\providecommand*{\natexlab}[1]{#1}
\providecommand*{\mciteSetBstSublistMode}[1]{}
\providecommand*{\mciteSetBstMaxWidthForm}[2]{}
\providecommand*{\mciteBstWouldAddEndPuncttrue}
  {\def\EndOfBibitem{\unskip.}}
\providecommand*{\mciteBstWouldAddEndPunctfalse}
  {\let\EndOfBibitem\relax}
\providecommand*{\mciteSetBstMidEndSepPunct}[3]{}
\providecommand*{\mciteSetBstSublistLabelBeginEnd}[3]{}
\providecommand*{\EndOfBibitem}{}
\mciteSetBstSublistMode{f}
\mciteSetBstMaxWidthForm{subitem}{(\alph{mcitesubitemcount})}
\mciteSetBstSublistLabelBeginEnd{\mcitemaxwidthsubitemform\space}
{\relax}{\relax}

\bibitem[Berg et~al.(2002)Berg, Tymoczko, and Stryer]{bertym:02}
Berg,~J.~M.; Tymoczko,~J.~L.; Stryer,~L. \emph{Biochemistry}, 5th ed.;
\newblock W. H. Freeman and Company, 2002\relax
\mciteBstWouldAddEndPuncttrue
\mciteSetBstMidEndSepPunct{\mcitedefaultmidpunct}
{\mcitedefaultendpunct}{\mcitedefaultseppunct}\relax
\EndOfBibitem
\bibitem[de~Planque and Killian(2003)]{depkil:03}
de~Planque,~M. R.~R.; Killian,~J.~A. Protein-lipid interactions studied with
  designed transmembrane peptides: role of hydrophobic matching and interfacial
  anchoring ({R}eview). \emph{Molecular Membrane Biology} \textbf{2003},
  \emph{20}, 271--284\relax
\mciteBstWouldAddEndPuncttrue
\mciteSetBstMidEndSepPunct{\mcitedefaultmidpunct}
{\mcitedefaultendpunct}{\mcitedefaultseppunct}\relax
\EndOfBibitem
\bibitem[May(2000)]{may:00}
May,~S. Theories on structural perturbations of lipid bilayers. \emph{Current
  Opinion in Colloid \& Interface Science} \textbf{2000}, \emph{5},
  244--249\relax
\mciteBstWouldAddEndPuncttrue
\mciteSetBstMidEndSepPunct{\mcitedefaultmidpunct}
{\mcitedefaultendpunct}{\mcitedefaultseppunct}\relax
\EndOfBibitem
\bibitem[Voth(2009)]{vot:09}
\emph{Coarse-Graining of Condensed Phase and Biomolecular Systems};
\newblock Voth,~G.~A., Ed.;
\newblock CRC Press, Boca Raton, 2009\relax
\mciteBstWouldAddEndPuncttrue
\mciteSetBstMidEndSepPunct{\mcitedefaultmidpunct}
{\mcitedefaultendpunct}{\mcitedefaultseppunct}\relax
\EndOfBibitem
\bibitem[M\"uller et~al.(2006)M\"uller, Katsov, and Schick]{muekat:06}
M\"uller,~M.; Katsov,~K.; Schick,~M. Biological and synthetic membranes: What
  can be learned from a coarse-grained description? \emph{Phys. Rep.}
  \textbf{2006}, \emph{434}, 113--176\relax
\mciteBstWouldAddEndPuncttrue
\mciteSetBstMidEndSepPunct{\mcitedefaultmidpunct}
{\mcitedefaultendpunct}{\mcitedefaultseppunct}\relax
\EndOfBibitem
\bibitem[Deserno(2009)]{des:09}
Deserno,~M. Mesoscopic Membrane Physics: Concepts, Simulations, and Selected
  Applications. \emph{Macromolecular Rapid Communications} \textbf{2009},
  \emph{30}, 752--771\relax
\mciteBstWouldAddEndPuncttrue
\mciteSetBstMidEndSepPunct{\mcitedefaultmidpunct}
{\mcitedefaultendpunct}{\mcitedefaultseppunct}\relax
\EndOfBibitem
\bibitem[Schmid(2009)]{sch:09}
Schmid,~F. Toy amphiphiles on the computer: What can we learn from generic
  models? \emph{Macromolecular Rapid Communications} \textbf{2009}, \emph{30},
  741--751\relax
\mciteBstWouldAddEndPuncttrue
\mciteSetBstMidEndSepPunct{\mcitedefaultmidpunct}
{\mcitedefaultendpunct}{\mcitedefaultseppunct}\relax
\EndOfBibitem
\bibitem[Marcelja(1976)]{mar:76}
Marcelja,~S. Lipid-mediated protein interactions in membranes. \emph{Biochimica
  et Biophysica Acta} \textbf{1976}, \emph{455}, 1--7\relax
\mciteBstWouldAddEndPuncttrue
\mciteSetBstMidEndSepPunct{\mcitedefaultmidpunct}
{\mcitedefaultendpunct}{\mcitedefaultseppunct}\relax
\EndOfBibitem
\bibitem[Dan et~al.(1993)Dan, Pincus, and Safran]{danpin:93}
Dan,~N.; Pincus,~P.; Safran,~S.~A. Membrane-Induced Interactions between
  Inclusions. \emph{Langmuir} \textbf{1993}, \emph{9}, 2768--2771\relax
\mciteBstWouldAddEndPuncttrue
\mciteSetBstMidEndSepPunct{\mcitedefaultmidpunct}
{\mcitedefaultendpunct}{\mcitedefaultseppunct}\relax
\EndOfBibitem
\bibitem[Aranda-Espinoza et~al.(1996)Aranda-Espinoza, Berman, Dan, Pincus, and
  Safran]{araber:96}
Aranda-Espinoza,~H.; Berman,~A.; Dan,~N.; Pincus,~P.; Safran,~S. Interaction
  between inclusions embedded in membranes. \emph{Biophysical Journal}
  \textbf{1996}, \emph{71}, 648\relax
\mciteBstWouldAddEndPuncttrue
\mciteSetBstMidEndSepPunct{\mcitedefaultmidpunct}
{\mcitedefaultendpunct}{\mcitedefaultseppunct}\relax
\EndOfBibitem
\bibitem[May and Ben-Shaul(2000)]{mayben:00}
May,~S.; Ben-Shaul,~A. A molecular model for lipid-mediated interaction between
  proteins in membranes. \emph{Physical Chemistry Chemical Physics}
  \textbf{2000}, \emph{2}, 4494--4502\relax
\mciteBstWouldAddEndPuncttrue
\mciteSetBstMidEndSepPunct{\mcitedefaultmidpunct}
{\mcitedefaultendpunct}{\mcitedefaultseppunct}\relax
\EndOfBibitem
\bibitem[Lag\"ue et~al.(2000)Lag\"ue, Zuckermann, and B.]{lagzuc:00}
Lag\"ue,~P.; Zuckermann,~M.~J.; B.,~R. Lipid-Mediated Interactions between
  Intrinsic Membrane Proteins: A Theoretical Study Based on Integral Equations.
  \emph{Biophysical Journal} \textbf{2000}, \emph{79}, 2867\relax
\mciteBstWouldAddEndPuncttrue
\mciteSetBstMidEndSepPunct{\mcitedefaultmidpunct}
{\mcitedefaultendpunct}{\mcitedefaultseppunct}\relax
\EndOfBibitem
\bibitem[Bohninc et~al.(2003)Bohninc, Kralj-Iglic, and May]{bohkra:03}
Bohninc,~K.; Kralj-Iglic,~V.; May,~S. Interaction between two cylindrical
  inclusions in a symmetric lipid bilayer. \emph{Journal of Chemical Physics}
  \textbf{2003}, \emph{119}, 7435--7444\relax
\mciteBstWouldAddEndPuncttrue
\mciteSetBstMidEndSepPunct{\mcitedefaultmidpunct}
{\mcitedefaultendpunct}{\mcitedefaultseppunct}\relax
\EndOfBibitem
\bibitem[Brannigan and Brown(2006)]{brabro:06}
Brannigan,~G.; Brown,~F. L.~H. A Consistent Model for Thermal Fluctuations and
  Protein-Induced Deformations in Lipid Bilayers. \emph{Biophysical Journal}
  \textbf{2006}, \emph{90}, 1501\relax
\mciteBstWouldAddEndPuncttrue
\mciteSetBstMidEndSepPunct{\mcitedefaultmidpunct}
{\mcitedefaultendpunct}{\mcitedefaultseppunct}\relax
\EndOfBibitem
\bibitem[Brannigan and Brown(2007)]{brabro:07}
Brannigan,~G.; Brown,~F. L.~H. Contributions of Gaussian Curvature and
  Nonconstant Lipid Volume to Protein Deformation of Lipid Bilayers.
  \emph{Biophysical Journal} \textbf{2007}, \emph{92}, 864--876\relax
\mciteBstWouldAddEndPuncttrue
\mciteSetBstMidEndSepPunct{\mcitedefaultmidpunct}
{\mcitedefaultendpunct}{\mcitedefaultseppunct}\relax
\EndOfBibitem
\bibitem[Schmidt et~al.(2008)Schmidt, Guigas, and Weiss]{schgui:08}
Schmidt,~U.; Guigas,~G.; Weiss,~M. Cluster Formation of Transmembrane Proteins
  Due to Hydrophobic Mismatching. \emph{Physical Review Letters} \textbf{2008},
  \emph{101}, 128104\relax
\mciteBstWouldAddEndPuncttrue
\mciteSetBstMidEndSepPunct{\mcitedefaultmidpunct}
{\mcitedefaultendpunct}{\mcitedefaultseppunct}\relax
\EndOfBibitem
\bibitem[de~Meyer et~al.(2008)de~Meyer, Venturoli, and Smit]{demven:08}
de~Meyer,~F. J.-M.; Venturoli,~M.; Smit,~B. Molecular Simulations of
  Lipid-Mediated Protein-Protein Interactions. \emph{Biophysical Journal}
  \textbf{2008}, \emph{95}, 1851--1865\relax
\mciteBstWouldAddEndPuncttrue
\mciteSetBstMidEndSepPunct{\mcitedefaultmidpunct}
{\mcitedefaultendpunct}{\mcitedefaultseppunct}\relax
\EndOfBibitem
\bibitem[West et~al.(2009)West, Brown, and Schmid]{wesbro:09}
West,~B.; Brown,~F. L.~H.; Schmid,~F. Membrane-Protein Interactions in a
  Generic Coarse-Grained Model for Lipid Bilayers. \emph{Biophysical Journal}
  \textbf{2009}, \emph{96}, 101--115\relax
\mciteBstWouldAddEndPuncttrue
\mciteSetBstMidEndSepPunct{\mcitedefaultmidpunct}
{\mcitedefaultendpunct}{\mcitedefaultseppunct}\relax
\EndOfBibitem
\bibitem[de~Meyer et~al.(2010)de~Meyer, Rodgers, Willems, and Smit]{demrod:10}
de~Meyer,~F. J.-M.; Rodgers,~J.~M.; Willems,~T.~F.; Smit,~B. Molecular
  Simulation of the Effect of Cholesterol on Lipid-Mediated Protein-Protein
  Interactions. \emph{Biophysical Journal} \textbf{2010}, \emph{99},
  3629--3638\relax
\mciteBstWouldAddEndPuncttrue
\mciteSetBstMidEndSepPunct{\mcitedefaultmidpunct}
{\mcitedefaultendpunct}{\mcitedefaultseppunct}\relax
\EndOfBibitem
\bibitem[Goulian(1996)]{gou:96}
Goulian,~M. Inclusions in membranes. \emph{Current Opinion in Colloid and
  Interface Science} \textbf{1996}, \emph{1}, 358--361\relax
\mciteBstWouldAddEndPuncttrue
\mciteSetBstMidEndSepPunct{\mcitedefaultmidpunct}
{\mcitedefaultendpunct}{\mcitedefaultseppunct}\relax
\EndOfBibitem
\bibitem[Weikl(2001)]{wei:01}
Weikl,~T.~R. Fluctuation-induced aggregation of rigid membrane inclusions.
  \emph{Europhysics Letters} \textbf{2001}, \emph{54}, 547\relax
\mciteBstWouldAddEndPuncttrue
\mciteSetBstMidEndSepPunct{\mcitedefaultmidpunct}
{\mcitedefaultendpunct}{\mcitedefaultseppunct}\relax
\EndOfBibitem
\bibitem[Reynwar et~al.(2007)Reynwar, Illya, Harmandaris, M{\"u}ller, Kremer,
  and Deserno]{reyill:07}
Reynwar,~B.~J.; Illya,~G.; Harmandaris,~V.~A.; M{\"u}ller,~M.~M.; Kremer,~K.;
  Deserno,~M. Aggregation and vesiculation of membrane proteins by
  curvature-mediated interactions. \emph{Nature} \textbf{2007}, \emph{447},
  461\relax
\mciteBstWouldAddEndPuncttrue
\mciteSetBstMidEndSepPunct{\mcitedefaultmidpunct}
{\mcitedefaultendpunct}{\mcitedefaultseppunct}\relax
\EndOfBibitem
\bibitem[Lenz and Schmid(2005)]{lensch:05}
Lenz,~O.; Schmid,~F. A simple computer model for liquid lipid bilayers.
  \emph{J. Mol. Liquids} \textbf{2005}, \emph{117}, 147--152\relax
\mciteBstWouldAddEndPuncttrue
\mciteSetBstMidEndSepPunct{\mcitedefaultmidpunct}
{\mcitedefaultendpunct}{\mcitedefaultseppunct}\relax
\EndOfBibitem
\bibitem[Schmid et~al.(2007)Schmid, D\"uchs, Lenz., and West]{schdue:07}
Schmid,~F.; D\"uchs,~D.; Lenz.,~O.; West,~B. A generic model for lipid
  monolayers, bilayers, and membranes. \emph{Computer Physics Communications}
  \textbf{2007}, \emph{177}, 168--171\relax
\mciteBstWouldAddEndPuncttrue
\mciteSetBstMidEndSepPunct{\mcitedefaultmidpunct}
{\mcitedefaultendpunct}{\mcitedefaultseppunct}\relax
\EndOfBibitem
\bibitem[Lenz and Schmid(2007)]{lensch:07}
Lenz,~O.; Schmid,~F. Structure of Symmetric and Asymmetric "Ripple" Phases in
  Lipid Bilayers. \emph{Physical Review Letters} \textbf{2007}, \emph{98},
  058104\relax
\mciteBstWouldAddEndPuncttrue
\mciteSetBstMidEndSepPunct{\mcitedefaultmidpunct}
{\mcitedefaultendpunct}{\mcitedefaultseppunct}\relax
\EndOfBibitem
\bibitem[West and Schmid(2010)]{wessch:10:2}
West,~B.; Schmid,~F. In \emph{NIC Symposium 2010}; M{\"u}nster,~G., Wolf,~D.,
  Kremer,~M., Eds.;
\newblock Forschungszentrum J{\"u}lich, 2010;
\newblock Chapter Membrane-Protein Interactions in Lipid Bilayers: Molecular
  Simulation versus Elastic Theory, pp 279--286\relax
\mciteBstWouldAddEndPuncttrue
\mciteSetBstMidEndSepPunct{\mcitedefaultmidpunct}
{\mcitedefaultendpunct}{\mcitedefaultseppunct}\relax
\EndOfBibitem
\bibitem[Neder et~al.(2010)Neder, West, Nielaba, and Schmid]{nedwes:10}
Neder,~J.; West,~B.; Nielaba,~P.; Schmid,~F. Coarse-Grained Simulations of
  Membranes under Tension. \emph{Journal of Chemical Physics} \textbf{2010},
  \emph{132}, 115101\relax
\mciteBstWouldAddEndPuncttrue
\mciteSetBstMidEndSepPunct{\mcitedefaultmidpunct}
{\mcitedefaultendpunct}{\mcitedefaultseppunct}\relax
\EndOfBibitem
\bibitem[Langel et~al.(2010)Langel, Cravatt, Gr{\"a}slund, von Heijne, Land,
  Niessen, and Zorko]{lancra:10}
Langel,~{\"U}.; Cravatt,~B.~F.; Gr{\"a}slund,~A.; von Heijne,~G.; Land,~T.;
  Niessen,~S.; Zorko,~M. \emph{Introduction to Peptides and Proteins};
\newblock CRC Press, 2010\relax
\mciteBstWouldAddEndPuncttrue
\mciteSetBstMidEndSepPunct{\mcitedefaultmidpunct}
{\mcitedefaultendpunct}{\mcitedefaultseppunct}\relax
\EndOfBibitem
\bibitem[Wimley and White(1996)]{wimwhi:96}
Wimley,~W.~C.; White,~S.~H. Experimentally determined hydrophobicity scale for
  proteins at membrane interfaces. \emph{Nature Structural Biology}
  \textbf{1996}, \emph{3}, 842--848\relax
\mciteBstWouldAddEndPuncttrue
\mciteSetBstMidEndSepPunct{\mcitedefaultmidpunct}
{\mcitedefaultendpunct}{\mcitedefaultseppunct}\relax
\EndOfBibitem
\bibitem[Bechinger(1996)]{bec:96}
Bechinger,~B. Towards Membrane Protein Design: {pH}-sensitive Topology of
  Histidine-containing Polypeptides. \emph{Journal of Molecular Biology}
  \textbf{1996}, \emph{263}, 768--775\relax
\mciteBstWouldAddEndPuncttrue
\mciteSetBstMidEndSepPunct{\mcitedefaultmidpunct}
{\mcitedefaultendpunct}{\mcitedefaultseppunct}\relax
\EndOfBibitem
\bibitem[Alberts et~al.(2002)Alberts, Johnson, Lewis, Raff, Roberts, and
  Walter]{albjoh:02}
Alberts,~B.; Johnson,~A.; Lewis,~J.; Raff,~M.; Roberts,~K.; Walter,~P.
  \emph{Molecular Biology of the Cell};
\newblock Garland Science, 2002\relax
\mciteBstWouldAddEndPuncttrue
\mciteSetBstMidEndSepPunct{\mcitedefaultmidpunct}
{\mcitedefaultendpunct}{\mcitedefaultseppunct}\relax
\EndOfBibitem
\bibitem[Venturoli et~al.(2005)Venturoli, Smit, and Sperotto]{vensmi:05}
Venturoli,~M.; Smit,~B.; Sperotto,~M.~M. Simulation Studies of Protein-Induced
  Bilayer Deformations, and Lipid-Induced Protein Tilting, on a Mesoscopic
  Model for Lipid Bilayers with Embedded Proteins. \emph{Biophysical Journal}
  \textbf{2005}, \emph{88}, 1778\relax
\mciteBstWouldAddEndPuncttrue
\mciteSetBstMidEndSepPunct{\mcitedefaultmidpunct}
{\mcitedefaultendpunct}{\mcitedefaultseppunct}\relax
\EndOfBibitem
\bibitem[Bohr(2009)]{boh:09}
\emph{Handbook of Molecular Biophysics};
\newblock Bohr,~H.~G., Ed.;
\newblock Wiley-VCH Verlag GmbH \& Co. KGaA, 2009\relax
\mciteBstWouldAddEndPuncttrue
\mciteSetBstMidEndSepPunct{\mcitedefaultmidpunct}
{\mcitedefaultendpunct}{\mcitedefaultseppunct}\relax
\EndOfBibitem
\bibitem[Scott and Coe(1983)]{scocoe:83}
Scott,~H.~L.; Coe,~T.~J. A Theoretical Study of Lipid-Protein Interations in
  Bilayers. \emph{Biophysical Journal} \textbf{1983}, \emph{42}, 219--224\relax
\mciteBstWouldAddEndPuncttrue
\mciteSetBstMidEndSepPunct{\mcitedefaultmidpunct}
{\mcitedefaultendpunct}{\mcitedefaultseppunct}\relax
\EndOfBibitem
\bibitem[Marsaglia(1972)]{mar:72}
Marsaglia,~G. Choosing a point from the surface of a sphere. \emph{Annals of
  mathematical statistics} \textbf{1972}, \emph{43}, 645--646\relax
\mciteBstWouldAddEndPuncttrue
\mciteSetBstMidEndSepPunct{\mcitedefaultmidpunct}
{\mcitedefaultendpunct}{\mcitedefaultseppunct}\relax
\EndOfBibitem
\bibitem[Schmid and Schick(1995)]{schsch:95}
Schmid,~F.; Schick,~M. Liquid phases of Langmuir monolayers. \emph{Journal of
  Chemistry Physics} \textbf{1995}, \emph{102}, 2080\relax
\mciteBstWouldAddEndPuncttrue
\mciteSetBstMidEndSepPunct{\mcitedefaultmidpunct}
{\mcitedefaultendpunct}{\mcitedefaultseppunct}\relax
\EndOfBibitem
\bibitem[Virnau and M\"uller(2004)]{virmue:04}
Virnau,~P.; M\"uller,~M. Calculation of free energy through successive umbrella
  sampling. \emph{Journal of Chemical Physics} \textbf{2004}, \emph{120},
  10925--10930\relax
\mciteBstWouldAddEndPuncttrue
\mciteSetBstMidEndSepPunct{\mcitedefaultmidpunct}
{\mcitedefaultendpunct}{\mcitedefaultseppunct}\relax
\EndOfBibitem
\bibitem[Schmidt et~al.(2009)Schmidt, Guigas, and Weiss]{schgui:09}
Schmidt,~U.; Guigas,~G.; Weiss,~M. Schmidt, {G}uigas and {W}eiss Reply.
  \emph{Physical Review Letters} \textbf{2009}, \emph{102}, 219802\relax
\mciteBstWouldAddEndPuncttrue
\mciteSetBstMidEndSepPunct{\mcitedefaultmidpunct}
{\mcitedefaultendpunct}{\mcitedefaultseppunct}\relax
\EndOfBibitem
\bibitem[Harroun et~al.(1999)Harroun, Heller, Weiss, Yang, and
  Huang]{harhel:99:1}
Harroun,~T.~A.; Heller,~W.~T.; Weiss,~T.~M.; Yang,~L.; Huang,~H.~W.
  Experimental Evidence for Hydrophobic Matching and Membrane-Mediated
  Interactions in Lipid Bilayers Containing Gramicidin. \emph{Biophysical
  Journal} \textbf{1999}, \emph{76}, 937--945\relax
\mciteBstWouldAddEndPuncttrue
\mciteSetBstMidEndSepPunct{\mcitedefaultmidpunct}
{\mcitedefaultendpunct}{\mcitedefaultseppunct}\relax
\EndOfBibitem
\bibitem[Sharpe et~al.(2002)Sharpe, Barber, Grant, Goodyear, and
  Marrow]{shabar:02}
Sharpe,~S.; Barber,~K.~R.; Grant,~C. W.~M.; Goodyear,~D.; Marrow,~M.~R.
  Organization of model helical peptides in lipid bilayers: {I}nsight into the
  behaviour of single-span protein transmembrane domains. \emph{Biophysical
  Journal} \textbf{2002}, \emph{83}, 345--358\relax
\mciteBstWouldAddEndPuncttrue
\mciteSetBstMidEndSepPunct{\mcitedefaultmidpunct}
{\mcitedefaultendpunct}{\mcitedefaultseppunct}\relax
\EndOfBibitem
\bibitem[Casuso et~al.(2010)Casuso, Sens, Rico, and Scheuring]{cassen:10}
Casuso,~I.; Sens,~P.; Rico,~F.; Scheuring,~S. Experimental Evidence for
  Membrane-Mediated Protein-Protein Interaction. \emph{Biophysical Journal}
  \textbf{2010}, \emph{99}, L47--L49\relax
\mciteBstWouldAddEndPuncttrue
\mciteSetBstMidEndSepPunct{\mcitedefaultmidpunct}
{\mcitedefaultendpunct}{\mcitedefaultseppunct}\relax
\EndOfBibitem
\bibitem[Humphrey et~al.(1996)Humphrey, Dalke, and Schulten]{humdal:96}
Humphrey,~W.; Dalke,~A.; Schulten,~K. VMD -- Visual Molecular Dynamics.
  \emph{Journal of Molecular Graphics} \textbf{1996}, \emph{14}, 33--38\relax
\mciteBstWouldAddEndPuncttrue
\mciteSetBstMidEndSepPunct{\mcitedefaultmidpunct}
{\mcitedefaultendpunct}{\mcitedefaultseppunct}\relax
\EndOfBibitem
\end{mcitethebibliography}

\end{document}